\documentclass[sigconf]{acmart}

\usepackage{color}

\usepackage{multirow}
\usepackage{balance}
\usepackage{enumitem}
\usepackage{url}
\usepackage{pdflscape}
\usepackage{afterpage}
\usepackage{changepage}
\usepackage{threeparttable}
\usepackage{wasysym}
\usepackage{caption}
\usepackage{tikz}
\usepackage{changepage}

\newcommand{\pie}[1]{%
\begin{tikzpicture}
 \draw (0,0) circle (0.8ex);\fill (0.8ex,0) arc (0:#1:0.8ex) -- (0,0) -- cycle;
\end{tikzpicture}%
}

\AtBeginDocument{%
  \providecommand\BibTeX{{%
    \normalfont B\kern-0.5em{\scshape i\kern-0.25em b}\kern-0.8em\TeX}}}

\setcopyright{none}
\copyrightyear{}
\acmYear{}
\acmDOI{}

\acmConference{}
\acmPrice{}
\acmISBN{}

\begin{document}

\title{A Survey of Privacy-Preserving Techniques for Encrypted Traffic Inspection over Network Middleboxes}

\author{Geong Sen Poh}
\email{geongsen.poh@trustwave.com}
\author{Dinil Mon Divakaran}
\email{dinil.divakaran@trustwave.com}
\author{Hoon Wei Lim}
\email{hoonwei.lim@trustwave.com}
\affiliation{
\institution{Trustwave}
}
\author{Jianting Ning}
\authornote{Work carried out when the authors were affiliated with the NUS-Singtel Cyber Security lab}
\email{jtning88@gmail.com}
\affiliation{
\institution{School of Information Systems \\ Singapore Management University}
\city{Singapore, 178902}
}
\author{Achintya Desai}
\email{achintya.desai@gmail.com}
\authornotemark[1]
\affiliation{
\institution{International Institute of Information Technology-Hyderabad}
\city{Hyderabad}
\state{India, 500032}
}

\begin{abstract}
Middleboxes in a computer network system inspect and analyse network traffic to detect malicious communications, monitor system performance and provide operational services. However, encrypted traffic, which has become increasingly prevalent, hinders the ability of middleboxes to perform such services. 
A common practice in addressing this issue is by employing a ``Man-in-the-Middle'' (MitM) approach, wherein an encrypted traffic flow between two end-points is interrupted, decrypted and analyzed by the middleboxes. 
The MitM approach is straightforward and is used by many organisations, but there are both practical and privacy concerns.
Practically, due to the cost of the MitM appliances and the latency incurred due to the encrypt-decrypt processes, enterprises continue to seek solutions that are less costly and less compute-intensive.
There has also been discussion on the many efforts required to configure MitM.
Besides, MitM violates end-to-end privacy guarantee, raising privacy concerns and potential issues on compliance especially with the rising awareness on user privacy.
Furthermore, some of the MitM implementations were found to be flawed. 
Consequently, new practical and privacy-preserving techniques that enable inspection over encrypted traffic were proposed. 

We systematically examine these techniques to compare their advantages, limitations and challenges. We categorise them into four main categories by defining a framework that consist of system architectures, use cases, trust and threat models. 
These are searchable encryption, access control, machine learning and trusted hardware. We first discuss the man-in-the-middle approach as a baseline, then discuss in details each of them, and provide an in-depth comparisons of their advantages and limitations. By doing so we describe practical constraints, advantages and pitfalls towards adopting the techniques. Following this, we give insights on the gaps between research work and practical implementation in the industries, which leads us to the discussion on the challenges and research directions. 
\end{abstract}




\maketitle

\section{Introduction}
\label{sec:Introduction}
Packet inspection and analysis have been used to detect, mitigate and block suspicious activities over home and enterprise networks. 
This is achieved by examining the headers and payloads of network traffic in real time.
The devices deployed for this purpose are known as middleboxes\footnote{A middlebox, also termed as a network appliance or a network function, is defined in RFC 3234 as {\em any intermediary box performing functions apart from normal, standard functions of an IP router on the data path between a source host and destination host}~\cite{CarpenterB02}. A similar definition is also provided in Part I of ETSI proposal on middlebox security protocol and enterprise transport security (ETS)~\cite{ETSIDraftI18}.}.
A middlebox (MB) provides various services and is indispensable in today's computer network infrastructure.
One of the main services include deploying MB for system and user security, for example, as personal and organizational firewall, intrusion detection and prevention system, parental filter, data exfiltration detection system, forensic analytic tool, as well as malware detection system.
In addition to security, it is also common to deploy an MB for performance and operational services.
These include service for proxies/caches as in content distribution network (CDN), WAN optimization, protocol acceleration, access control, billing and usage monitoring, and network address translations. 
Compliance service is also deployed as an MB, to fulfill obligations such as the need to support lawful interception and control of illicit content and privacy. 

\begin{figure}[!ht]
	\centering
	\includegraphics[scale=0.30]{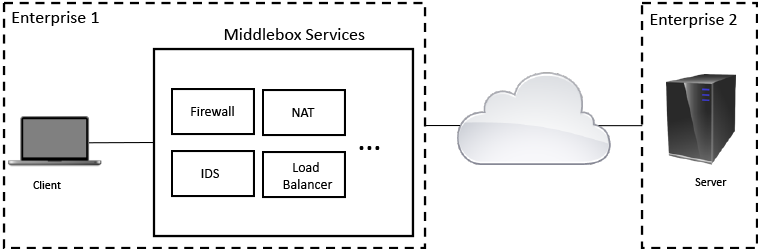}
	\caption{Traditional Model-I [Client-Oriented]: Client connects to middlebox service for in-bound and out-bound network traffic inspections}\label{fig:MBArch1}
\end{figure}

\begin{figure}[!ht]
	\centering
	\includegraphics[scale=0.30]{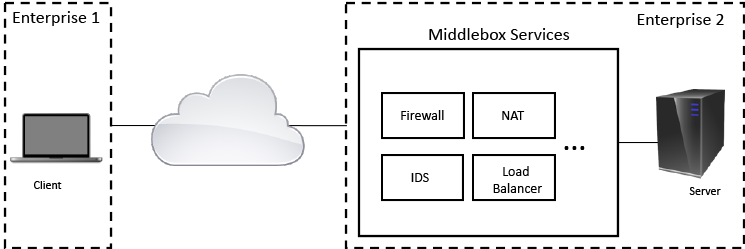}
	\caption{Traditional Model-II [Server-Oriented]: Server connects to middlebox service for in-bound and out-bound network traffic inspections}\label{fig:MBArch2}
\end{figure}

\begin{figure}
	\centering
	\includegraphics[scale=0.30]{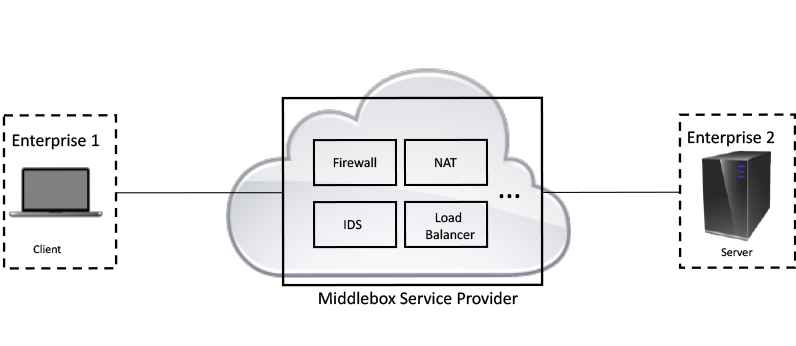}
	\caption{Outsourced Model: Client or host server subscribes to cloud-based middlebox service}\label{fig:OutMBArch}
\end{figure}

Traditionally, MBs are deployed at the premises of enterprise or customer networks. These are known as `on-premise' solutions (Figures~\ref{fig:MBArch1} and~\ref{fig:MBArch2}). With the advent of Network Function Virtualisation (NFV) \cite{HanGJL15}, the dependence on specialised and expensive hardware for deployment of MBs is also being challenged, and there is a clear shift towards deployment of software-based middlebox functions. In other words, we have started to see the outsourcing of MBs to the cloud infrastructure from the common on-premise deployment model~\cite{MB-outsourcing-2010,LanSPRL16} (Figure~\ref{fig:OutMBArch}). While cloud infrastructures provide more flexibility and dynamic scalability than a hardware-based appliance, they also come with new challenges, especially in ensuring security and privacy of these systems~\cite{BhargavanBDFO18,ETSIDraftI18}.

In order to enable services mentioned above in an effective manner, middleboxes often perform network inspection and analysis using a well-established technique known as deep packet inspection (DPI)\footnote{According to the International Telecommunication Union recommendation ITU-T Y2770~\cite{ITUT12}, DPI is the {\em analysis, according to the layered protocol architecture OSI-BRM [specified in ITU-T X.200], of payload and/or packet properties [listed in clause 3.2.11] deeper than protocol layer 2, 3 or 4 header information, and other packet properties, in order to identify the application unambiguously.}}. 
DPI implements in-depth inspection on the headers and payloads of network packets, in contrast to traditional packet filtering that inspect only packet headers. DPI can be stateful, with different useful states stored during packet processing, such as flow characteristics, application status, etc~\cite{ITUT12}.
However, currently $87\%-90\%$ of the network traffic are encrypted using TLS~\cite{Meeker19,Cisco18}. According to Google transparancy report, as of November 2020, $81\%-98\%$ of the traffic using Chrome platform across different operating systems are HTTPS traffic~\cite{GoogleTR20}. Existing DPI techniques that are only capable of inspecting plain packet properties would be of limited usage and critically impact on the various network services as was detailed by Carnavalet and Van Oorschot in~\cite{Carnavalet20}. 
 This means an MB must devise mechanisms capable of analysing encrypted traffic in a manner that balances the requirements of privacy, utility and performance.

\subsection{Industry Practices and New Approaches}
\label{subsec:IndustryPractices}
Two common techniques widely used in the industry (in particular to deploy MBs at enterprises~\cite{Broadcom20}) to inspect encrypted traffic is (1) the split-TLS technique, which is also known as a man-in-the-middle (MitM) approach, and (2) key sharing and delegation.

\paragraph{\bf MitM} In MitM, instead of establishing an end-to-end TLS session between the client and the server, the client establishes a session with the middlebox.
By doing so, encrypted traffic originating from the client can be decrypted, inspected and analysed by the MB.
The MB re-encrypts and forwards the data to the server on behalf of the client via a second, new TLS session between the MB and the server.
This provides a practical solution that can be deployed without requiring any changes to the TLS protocol, but it does require a client to install the MB's root certificate.
The root certificate enables the MB to present itself as the server (i.e. the destination endpoint) to the client by copying and signing a new certificate based on the credentials of the server.
In effect, the MB impersonates the server.
This deployment is secure as long as the root certificate is securely stored, up-to-date TLS implementation are used, and a configurable policy engine that enables an administrator to set a whitelist\footnote{A whitelist contains sites where the middleboxes would not inspect and merely forward the encrypted traffic. For example, encrypted traffic to online banking sites, and credit card transactions.} is provided.
Unfortunately, some of the deployments have been shown to be insecure due to weaknesses in their implementation of the underlying protocols, such as allowing deprecated cipher suites, as discussed by Jarmoc~\cite{Jarmoc12}, Carnavalet and Mannan~\cite{CarnavaletM16}, Durumeric {\em et\ al.}~\cite{DurumericMSBSBB17} and Waked{\em et\ al.}~\cite{WakedMY18}.
Also, based on the surveys conducted by Sherry {\em et\ al.}~\cite{Sherry16}, a large network with heterogeneous network devices would require many experienced administrators to administer them.
This poses another privacy concern when MitM approach is used, since many of these devices may have access to the decrypted data.
It would be difficult to configure which are the devices that should have access and also to trace the network traffic.

\paragraph{\bf Key Sharing and Delegation}
In this approach, enterprises share their certificates, or private keys, and clients share session keys with the middleboxes. They are supported by industries for practical use. For example, a server sharing its private key with a CDN, or the servers share static keys with middleboxes, as detailed in the Enterprise Transport Security (ETS) standardised by the ETSI~\cite{Carnavalet20}. Issues with some of these approaches include not supporting Perfect Forward secrecy (PFS), and hence will not be compatible with TLS 1.3. Detailed discussion on the various approaches and issues can be found in~\cite{Carnavalet20}.

Both the MitM and key sharing approaches violate the end-to-end encryption and data privacy guarantee of the supposedly secure two-party communications, and can be costly and difficult to implement.
There is also the possibility that an MB stores the decrypted data, which
opens up another set of possibilities for attack and data leak.
It means the MBs and the administrator must be fully trusted in such a setting.
Due to these, US-Cert has recently issued an alert stating that using HTTPS interception weakens TLS security~\cite{CISA17} and the National Security Agency (NSA) has also issued an advisory on potential insider threats on the use of MitM~\cite{NSA19}.

\paragraph{\bf Privacy-Preserving Approaches} From the perspective of privacy, a survey on user acceptance on encrypted traffic inspection (i.e.\ TLS) by middleboxes was conducted by Ruoti {\em et\ al.}~\cite{RuotiAMZS18}.\ $1976$ participants was surveyed. $75.8\%$ express concerns about privacy and identity theft by hackers, while $70.9\%$ are concerns about government surveillance.
Furthermore, using MitM and key sharing as a tool to inspect network traffic may also violate privacy regulation such as the recently announced The EU General Data Protection Regulation (GDPR).
For example, explicit consent from a consumer may be required for a third-party (e.g.\ the network security service provider that supplies and manages the MBs) to access the data, and $83.2\%$ of participants in Ruoti {\em et\ al.}'s survey indicated that they should first be notified or consent. 

It is due to the above concerns new approaches are proposed by the research community.
Many proposals focus on the enterprise environment as discussed above (e.g. the client-oriented and server-oriented setting in Figures~\ref{fig:MBArch1} and~\ref{fig:MBArch2} respectively).
One new approach is to perform encrypted traffic analysis (ETA) based on machine learning techniques.
A comprehensive study can be found in the ENISA recent report~\cite{ETAENISA19}.
The report surveyed six use cases that machine learning techniques are largely effective, namely application identification, network analytics, user information identification, detection of encrypted malware, fingerprinting and DNS tunnelling detection.
Machine learning techniques preserve privacy as they do not inspect the (encrypted) payloads. However, it was noted that such techniques cannot offer the same level of inspection as in normal monitoring of unencrypted traffic.

On the other hand, another new approach is to devise privacy-preserving techniques to inspect encrypted payloads without the pitfalls of MitM.
However, the proposed schemes require either active participation of both the client and server endpoints, or introduce accountability on both endpoints to allow visibility on all the MBs that sit between the client and the server.
In practice, based on our interaction with the industries, it is not feasible to request all endpoints to install or adhere to such an architectural requirement, as was also pointed out by Carnavalet and Van Oorschot~\cite{Carnavalet20}.
For example, detection should be performed in a way that is agnostic to the destination.
A client machine, as the source endpoint, visits web application providers such as Google, Facebook, Amazon or Instagram.
It would be a major undertaking for the providers to have to deploy similar solution of that of clients from different enterprises visiting their sites.
Also, how would the providers trust the enterprises that request to establish connection, and what is the incentive in going through the potentially expensive customised setup cost?
We believe this is one of the obstacles in adopting existing privacy-preserving proposals for actual deployment in practice.
We discuss schemes using both approaches in details in Section~\ref{sec:Techniques}.

\subsection{Outsourced MB Services}
The difficulty of administering the myriad devices~\cite{Sherry16} has fuelled the development of new middlebox services that departs from the traditional on-premise enterprise middlebox setting.
The emerging trend is to outsource MB functionality to cloud-based services, commonly known as MB as a services.
Figure~\ref{fig:OutMBArch} illustrates a general architecture of outsourced MBs.
It relieves an enterprise from needing to purchase related hardware, install (both software and hardware), configure, operate and maintain middleboxes. Outsourcing of MBs therefore reduces, or virtually removes, operational costs, and thereby enables the enterprise to focus on its key business.
However, outsourcing MBs to a third-party provider exacerbates privacy concern when compared to the on-premise enterprise setting, since now, data that are normally inspected internally, has to be routed to the provider for processing.
Therefore, many research works have proposed to secure MB services in the cloud, including privacy-preserving inspections on encrypted traffic.
In addition to this, cloud-based MB must provide low-latency operations since traffic has to be rerouted to the cloud.
An example solution that enables privacy-preserving inspections with low-latency is Embark, proposed by Lan {\em et\ al.}~\cite{LanSPRL16}.
Low-latency is achieved through an architecture known as Appliance for Outsourcing MBs (APLOMB) introduced in~\cite{SherryHSKRS12}.
The general idea is to create a tokenised encrypted traffic along the TLS traffic.
Searchable encryption method is used to match the tokenised encrypted traffic with the encrypted rulesets at the MB, all of which are setup under the APLOMB architecture.


In addition, many more recent outsourced-based MB proposals develop techniques that are based on secure enclave (i.e. Intel SGX).
These include, for example, SafeBricks~\cite{PoddarLPR18}, ShieldBox~\cite{TrachKGABF18} and SPlitBox~\cite{AsgharMSCKM16}.
The general idea of these techniques is to perform packet inspections in the secure enclave so that the cloud provider does not learn the content of the encrypted packets.
We discuss in more details these proposals in Section~\ref{subsec:TH}.



\subsection{Overview of Our Study}
In essence, one is looking for an optimal solution between two extremes, (1) of not being able to perform in-depth inspection due to the traffic being encrypted; and (2) inspection of decrypted traffic using MitM and key sharing approach.
Many solutions have been proposed in the recent years to address this issue.
We categorise them into four categories based on the techniques they used. 
We discuss these techniques in Section~\ref{sec:Techniques}. We also include the MitM approach as a fifth category in the beginning of the discussion for comparisons.
We believe the different designs and properties of the growing list of proposals should be studied in order to clearly characterise their advantages and limitations.
Our study provides insights on security assurance and performance over different use case scenarios, as well as appeals of the techniques to real-world deployment:


\begin{itemize} [leftmargin=*]


\item We define a general architecture, identify use cases and their constraints (Section~\ref{sec:ArchNusecase}).
We define three system models under the prescribed architecture that illustrates the flow of network traffic through the MBs.
We term them client-oriented, server-oriented and client-server accountable settings.
Our models provide a more fine-grained categorisation and enhancement on the one-sided and two-sided definition in the ETSI standard~\cite{ETSIDraftI18,ETSIMSP20}.

\item We define a trust model encompassing the different actors in a network system that contains MBs (Section~\ref{sec:TrustNPractice}).
It provides a foundation on which security model can be defined clearly.
We state trust assumptions based on the architecture and settings that we defined, and define threats and security requirements based on these assumptions.

\item We classify existing privacy-preserving techniques into two types, {\em passive} and {\em active}.
Here, {\em passive} inspection encompasses techniques that analyse encrypted traffic without decryption or, modify the traffic of the underlying protocol.
{\em Active} inspection analyses the traffic by decryption or modification of the underlying protocol.
These techniques are categorised as {\em access control}, {\em searchable encryption}, {\em machine learning} and {\em trusted hardware}, as well as the MitM approach for comparison (Section~\ref{sec:Techniques}).
We compare the main features, advantages and pitfalls of each category. 

\item We identify the research challenges and discuss the potential research directions (Section~\ref{sec:ChallengesNResearchDirections}).
We examine the current security model and analyse the existing attacks.
This leads us to believe that some of the existing proposals require further improvements for industry adoptions. For example, a protocol that utilises searchable encryption mechanisms will need to be analysed against well-established notion of information leakages in the field.
We give insights, especially on the reason that existing MitM approaches, with combination of a configurable policy engine, are still preferred in the industries.

\end{itemize}
As a side note, for practical and efficient deployment with optimal inspection capability, it seems decryption of the encrypted payload is one of the better approaches.
What remains to be studied is how much encrypted data should be decrypted and revealed to the MBs in order to preserve privacy.
It is, we believe, based on this fact that a new MB security protocol (MSP) was proposed and is undergoing standardisation efforts~\cite{ETSIDraftII18}.
It may also be the case that trusted hardware approaches, such as using Intel SGX, are being developed where encrypted traffics are decrypted in a way protected by the hardware so that MBs have no access to the decrypted data.
The techniques and challenges are discussed in details in Section~\ref{sec:Techniques} and Section~\ref{sec:ChallengesNResearchDirections}. 

A short survey related to the issue that we studied was provided by Wang {\em et\ al.} in~\cite{WangYCR18}. 
It studied the various mechanisms for secure outsourcing of MBs in a general manner, focusing solely on cloud-based MB.
Our work complements theirs in a way that we provide broader investigation that also covers non-outsourced specific literature, hence our definition of different system models and use cases.
The European Union Agency for Cybersecurity (ENISA) also published a survey on encrypted traffic analysis~\cite{ETAENISA19}. The survey describes in detail in particular $6$ key use cases (i.e. application identification, network analytics, user information identification, detection of encrypted malware, file/device/website/location fingerprinting and DNS tunnelling detection) and techniques based on machine learning.
Most recently, Carnavalet and Van Oorschot~\cite{Carnavalet20} presented a comprehensive survey on TLS interception mechanisms and motivations, focusing on practical considerations between the use cases and incentives of the stakeholders.
They listed $19$ use cases where access to unencrypted traffic is crucial, in order to understand the motivation of the various proposals in inspecting encrypted traffic.
One of the key insights from their survey is the identification of gaps between the proposed mechanisms, use cases and incentives.
Extensive details on the various MitM and key sharing techniques were studied and compared.
Here, our focus is on privacy-preserving techniques with compilations, categorisation of the state-of-the-arts, in-depth examinations and comparisons of these various techniques, which complement these recent surveys. 

\section{System Models and Use Cases }
\label{sec:ArchNusecase}
We now describe the underlying models and use cases. 
The different models that we discuss here are based on the setting provided by Sherry {\em et\ al.} (BlindBox)~\cite{SherryLPR15}, Canard {\em et\ al.} (BlindIDS)~\cite{CarnavaletM16}, Bhargavan {\em et\ al.}~\cite{BhargavanBDFO18} and ETSI standard I~\cite{ETSIDraftI18,ETSIMSP20}.
In order to provide practical insights, we also present use case scenarios for each models.
Figure~\ref{fig:MBArch1},~\ref{fig:MBArch2} and~\ref{fig:OutMBArch} show a general high-level architecture of MBs operating in a computer network environment.
We describe existing and potential use cases based on these architectures.
We remark that it is common for MBs to act directly to the passing traffic in an {\em in-band} middlebox setting where one or more middleboxes are placed in-line between the client and the server. Common use cases for operational purposes include content delivery networks (CDNs), assess control, billing and usage monitoring, asset tracking, name or tag resolution and operations control. It is also possible in certain use cases that MBs store the traffic and analyse them in an {\em out-of-band} setting. Cyber security use cases include network firewalls, application firewalls, intrusion detection system (IDS) and intrusion prevention system (IPS). It is also used for compliance obligations, for example availability/resilience, emergency and public safety communication, data retention, identity management, cyber security, content control, personal data and privacy ~\cite{ETSIDraftI18,ETSIMSP20}.

%

\subsection{Client-Oriented MBs}
We first describe a typical client-oriented setting that is common in an enterprise network.
Figure~\ref{fig:MBArch1} illustrates a high-level model of the setting, where the goal of the middleboxes is to protect the clilent(s), and therefore are often placed closer to the client(s). 
Here, a client connects to a server using a secure channel.
The encrypted traffic flows, both inbound and outbound, are routed through a host of MBs that inspect the traffic. MBs receive in-bound and out-bound traffic from multiple clients and server endpoints.
It normally involves installing specific software, or modification on the connections at the client device in order to enable the MBs to perform the inspections.
\bigbreak
\noindent{\em Use Cases.} Common use cases include firewalls, advertisement blocking, personal data protection and anonymisation of information~\cite{deutschetel2016zscaler, tmobile-secureweb2019}. An example is, protection of user's privacy while enabling encrypted traffic inspection in a computer network of an organisation (e.g. a university, a financial institution or a telecommunication company) or by an outsourced middlebox service.
Sherry {\em et\ al.}~\cite{SherryLPR15} described a scenario where students using the university network install the proposed solution to preserve privacy of their data yet allowing the system (e.g., a security solution) to inspect the encrypted traffic.
Another use case, of parental filter, where a user subscribes the service from an ISP (internet service provider) is also presented.
These two examples prioritises inspection of outbound traffic from the client.
Bhargavan {\em et\ al.}~\cite{BhargavanBDFO18} also stated that firewalls are deployed in companies and educational institutions to protect computers from malware on both inbound and outbound traffic, for example. 
The most relevant deployment scenario for enterprises is the client-oriented middlebox. In this scenario, an enterprise deploys an MB appliance at its premise, at the perimeter of the network. The MB inspects all incoming and outgoing traffic (except those whitelisted) from/to the clients, using the previously mentioned MitM approach. Since enterprises often enforce policies on employee machines, the installation of root certificates that allow MBs to decrypt and re-encrypt the traffic is not a practical issue.


\subsection{Server-Oriented MBs}
A server-oriented MB is also common in an enterprise network (Figure~\ref{fig:MBArch2}). This is to secure the servers hosted in an enterprise, for example web servers.
A server in an enterprise establishes secure channels with requesting clients that are often outside the enterprise network, in particular, the Internet. 
The encrypted traffic to/from the server endpoint are routed through a host of MBs at the server-side (or outsourced to a third-party service provider) that inspect the encrypted traffic.
The main technical difference in this setting compared to the previous one is that connections are all requested by clients, and in general the purpose is to secure one specific server.
Server-oriented MBs are developed to protect servers, such as web servers, database servers, and so on. A typical example is a web application firewall that inspects incoming HTTP requests to a web server. As the web server itself has moved to the cloud since a few years now, having server-oriented MBs in the cloud is not uncommon these days~\cite{Akamai-web-sec}. 
\bigbreak
\noindent{\em Use Cases.} A common scenario is where content delivery network (CDN) serves TLS traffic on behalf of customer websites.
In this case, the emphasis is on inspecting the many inbound traffic flows to the websites.
Another use case is provided in the ETSI standard~\cite{ETSIDraftI18,ETSIMSP20}.
It involves a data center that would require inspection on encrypted traffic from various internal networks.
This means MBs between these networks must be able to communicate and inspect the underlying encrypted traffic.

\section{Trust Model}
\label{sec:TrustNPractice}
{\bf At least one of the two endpoints must be honest.} While deploying a traditional network security solution, such as an intrusion detection system (IDS), it is assumed that either one of the endpoints must be honest~\cite{SherryLPR15,YuanWLW16}.
Otherwise two malicious endpoints can agree on a secret key and encrypt the traffic using well-established encryption schemes.
Canard {\em et\ al.}~\cite{CarnavaletM16} described a scenario, whereby an infected bot connects to the remote command and control server using an encrypted channel.
Inspection will be impossible when the traffic flows are encrypted.
Similar assumption is made in the case of MBs providing parental filtering and data exfiltration detection~\cite{SherryLPR15}.
In parental filter, it is assumed innocent child would not replace network protocol stack or install tunneling software.
Data exfiltration is assumed to occur due to accidental transmission of sensitive data.
In other words, the case of a user (or an adversary that gained control on the device) intentionally sabotaging a device to circumvent existing network connection is not considered in some of the proposals~\cite{SherryLPR15,LanSPRL16,CanardDKPS17,NingPLCC19,NingHPXLWD20}.
\bigbreak
\noindent {\bf What if both endpoints are malicious?} Given the above description, we may argue the common assumption is that either the client or the server must be honest.
Nevertheless, one cannot discount an attack that intentionally modify a user's device, for example, to extract and send sensitive information to a malicious server.
In fact, a scheme that assumes such a scenario, where both the user and the server are malicious is proposed by Goltzsche {\em et\ al.}~\cite{GoltzscheRN+18}.
Their proposed solution utilizes secure enclave to prevent a malicious client from manipulating the TLS traffic.
\bigbreak
\noindent From these contrasting assumptions, in the following we define in clearer details, the trust assumptions under the three models and use cases that we have discussed.

\subsection{Assumptions}
\noindent {\bf Entities.} Three main entities are involved in a typical setting.
These are a {\em client}, a {\em MB provider} and a {\em server}.
In some of the schemes, a fourth entity may be involved.
For example, in the outsourced MB setting, a {\em cloud service provider} hosts the middleboxes.
In other settings, an {\em appliance} acts as a trusted party in an enterprise for registration or traffic encryption, or a {\em rule generator} sharing/encrypting the rulesets for the middleboxes.
An entity can be {\em honest}, {\em semi-honest}, or {\em malicious}.
By {\em honest}, we mean that an entity follows the protocol honestly and adheres to all the security requirements and goals of the protocol.
On the other hand, a {\em semi-honest} entity follows the protocol honestly, but also `listens' to the communications in order to try to extract or learn the content of the messages from the communications.
A semi-honest entity is also known as an {\em honest-but-curious} entity in the literature.
A {\em malicious} entity is an entity that in addition to listening to the communications, has the capability to modify the communications in order to, for instance, impersonate one of the legitimate entities and learn the underlying messages being transmitted, or assume control of the legitimate entity.
We note that it is possible that a client is infected by bot in an enterprise network (hence becoming a {\em malicious} entity) but the protocols are not affected. That is to say, the traffic flows still route through the enterprise network and the middlebox.
\bigbreak
\noindent{\bf Trust.} Table~\ref{tbl:YES} lists the trust assumptions considered by existing schemes.
A common setting is that the middleboxes are either honest (H) or semi-honest (S), while either the client or the server is honest\footnote{We do not consider the case of semi-honest client/server since the original intend of the communication is for both of them to send/receive messages, and hence they learn the underlying content anyway.}.
This is the case for schemes using four techniques that we discussed in Section~\ref{sec:Techniques}. These techniques are Man-in-the-Middle (MitM), searchable encryption (SE), access control (AC) and machine learning (ML) techniques, as shown in row 1--4 in the table.
An exception is a scheme called EndBox~\cite{GoltzscheRN+18} (row $5$), where both the client and the server can be malicious, albeit the MB is assumed honest.
This is made possible by deploying trusted hardware (TH) such as secure enclave (i.e. Intel SGX) at the client's device.
In this way, the attacker is not able to access the scheme and modify the executions of the code that are carried out inside the enclave.
Obviously if all three entities are malicious then there is nothing to be protected. 
Secure enclave was also used in more recently proposed schemes in the outsourced MB scenario. 
Examples include SafeBrick~\cite{PoddarLPR18}, ShieldBox~\cite{TrachKGABF18} and SGX-Box~\cite{HanKHH17}.
Here, the service provider hosting the MBs can be malicious since executions of the MB functionalities, including packet inspection are performed in the secure enclave.
It means a malicious service provider should not be able to learn any information from or modify the executions of the enclave.

\begin{table*}[!ht]
\centering
\caption{Trust Assumptions of Existing Schemes and Solutions\label{tbl:YES}}{
\scriptsize
\begin{threeparttable}
\begin{tabular}{@{}|l|c|c|c|c|c|c|p{8cm}|}
\hline
Scheme & Tech. (Sec.~\ref{sec:Techniques}) & Client & MB & CSP & Server & Model & Remarks\\
       &  & & & & & &\\\hline
Commercial solutions and & MitM & H/M & H & n/a & M/H & C-O & Preserve privacy through a whitelist policy engine (e.g. traffic for online banking is  \\ 
 specific techniques~\cite{Carnavalet20} &  &  &  &  & & S-O & not decrypted). \\\hline
BlindBox~\cite{SherryLPR15}, SPABox~\cite{FanGRCQ17}, &  &  &  &  &  &  & \\
BlindIDS~\cite{CanardDKPS17}, Embark~\cite{LanSPRL16}, & SE & H/M & S & n/a & M/H & C-O & Introduce (semi-)honest rule generator$^*$. No rule generator for SplitBox, but requires \\
Yuan {\em et\ al.}~\cite{YuanWLW16}, SplitBox~\cite{AsgharMSCKM16} &  &  & & & & S-O & multiple cloud providers. \\
PrivDPI~\cite{NingPLCC19}, Pine~\cite{NingHPXLWD20} &  &  & & & & & \\\hline
mcTLS~\cite{NaylorSVLBLPRS15}, mbTLS~\cite{NaylorLGKS17}, maTLS~\cite{LeeSLCCCK19} & AC & H/M & -- & n/a & M/H & C-S & MBs visible to both endpoints.\\
MSP~\cite{ETSIDraftII18}, Bhargavan {\em et\ al.}~\cite{BhargavanBDFO18} &  &  &  &  & &  & \\\hline
Anderson {\em et\ al.}~\cite{AndersonM16, AndersonM17, AndersonPM18} & ML & H/M & S & n/a & M/H & C-O &  Analyze metadata of the encrypted traffic. \\
Yamada {\em et\ al.}~\cite{YamadaMTSP07} &  &  & & & & S-O & \\\hline
EndBox~\cite{GoltzscheRN+18} & TH & M & H & H & -- & C-O & Deploy secure enclave as MB on client.\\\hline
SafeBricks~\cite{PoddarLPR18}$^+$, ShieldBox~\cite{TrachKGABF18}$^+$ & TH & H & M & M & M & C-O & MB hosted by service provider can be malicious. Secure enclave in MB performs \\
SGX-Box~\cite{HanKHH17}, LightBox~\cite{DuanYW17}  &  & M & M & M & H & S-O & inspection instead. \\\hline
\end{tabular}
\begin{tablenotes}\footnotesize
\item[] H: Honest; S: Semi-honest; M: Malicious; --: Either H, S or M. C-O: Client-Oriented; S-O: Server-Oriented; C-S: Client-Server Accountable; MB: middlebox; CSP: Cloud Service Provider.
\item[] For technique, MitM: Man-in-the-Middle, SE: Searchable Encryption, AC: Access control, ML: Machine Learning, TH: Trusted Hardware. Each technique is discussed in details in Section~\ref{sec:Techniques}.
\item[] *: Lan {\em et\ al.} (SafeBrick) provides three different rule settings: (1) Both client and MB know the rules; (2) Only client should know the rules. In this case client encrypts the rulesets and passes them to the MB; (3) Only MB should know the rules. In this case a trusted rule generator is required to generate signatures on the rules so that the MB cannot simply generate rules to match arbitrary data from the encrypted traffic.
\item[] +: Solution for outsourced MB services. Cloud provider can be malicious but network traffic is protected under a secure enclave implemented in MB.
\end{tablenotes}
\end{threeparttable}}
\end{table*}

In terms of system models, many of the schemes cater for client-oriented and server-oriented setup (C-O and S-O) as discussed in Section~\ref{sec:ArchNusecase}.
These schemes can be straightforwardly deploy on either the client enterprise network, or at the server (e.g. web application provider).
Schemes based on the searchable encryption (SE) technique such as BlindBox~\cite{SherryLPR15} (c.f. second row of Table~\ref{tbl:YES}), however, further assumes the existence of a (semi-)honest rule generator that provides rulesets to MB.
There are also schemes that cater for client-server accountability (c.f. row 3), in which both endpoints have visibility of the MBs deployed and are able to authenticate these MBs.
On the other hand, EndBox~\cite{GoltzscheRN+18} and SplitBox~\cite{AsgharMSCKM16} focus on client-oriented model, with EndBox providing stronger security assurance where client can be malicious.
In summary, there are two main trust assumptions considered by the existing schemes:

\begin{itemize}
    \item {\bf Trust Assumption (TA) I:} Middlebox (or MB provider) is (semi-)honest and one of the endpoints (i.e. client or server) must be honest.
    Schemes based on this assumption in general deploy MitM, SE, AC, ML techniques.
    \item {\bf Trust Assumption (TA) II:} Cloud service provider and one of the endpoints (i.e. client or server) can be malicious.
    Alternatively, both endpoints can be malicious, but the MB provider is honest.
    Schemes based on this assumption in general deploy TH techniques.
\end{itemize}
The discussed schemes cover only assumptions that relate to the perceived use cases and scenarios.
Certainly, there are obvious trust combinations that are not feasible, for example:
\begin{itemize}
    \item In the case whereby the client and the server are honest then MBs would be redundant.
    The client and the server communicates directly through encrypted traffic.
    Their main security requirements would be to prevent outsiders from listening or modifying their communication.
    \item It is obvious that if all three entities are malicious then there is nothing to protect from.
\end{itemize}
New trust assumptions may be required due to new requirements and scenarios that were not addressed by existing studies.
Here we give an example, on the possibility of collusion between the service provider hosting the MBs and one of the endpoints, when we consider the outsourced MB scenario.

\paragraph{\bf Collusion.}
 Note that if MB is malicious (or semi-honest), and either the client or the server is malicious (or semi-honest), the network traffic is exposed to the MB and data can be modified or viewed if the {\em two entities collude}.
This is the case because either the client or the server can share the session key with the MB.
If there are many MBs, and some of them are malicious and others are honest, then we may treat this as the scenario of honest client, honest server and malicious MB, by assuming the honest MBs holding the role of clients or servers.
At first glance, this may not seem to be practical since it does not make business sense for a service provider to want to learn information from the subscriber by colluding with the other endpoint.
However, it can be an unscrupulous employee of the service provider that collude, or the service provider and one of the endpoint being directed to extract information under a government surveillance program.

%
%
%
%
%

\subsection{Security Requirement}
\label{subsec:secReq}
The main goal of schemes advocating privacy-preserving inspection of encrypted traffic is to protect data privacy of the encrypted network traffic, while maintaining similar utility of inspection on plain data traffic.
The main security requirement is thus to ensure {\em data privacy}, in such a way that only the endpoints know the underlying message of the encrypted traffic while the entity that performs the inspection does not learn any information {\em it is not allowed to learn}.
Depending on schemes, the information that is learned by the inspecting entity can vary.
For instance, in a Man-in-the-Middle (MiTM) approach that decrypts the network traffic is able to learn all the information if it chooses to.


\section{Techniques}
\label{sec:Techniques}
We survey existing state-of-the-art techniques for privacy-preserving inspection of encrypted network traffic, which were briefly stated in the previous section.

We first define two types of inspections used by these techniques.


\begin{itemize} [leftmargin=*]

\item {\em Passive Inspection.}
A scheme provides passive inspection if it (1) does not modify the underlying protocol (i.e. SSL/TLS) and (2) does not decrypt the encrypted traffics in order to perform inspection.

\item {\em Active Inspection.}
A scheme provides active inspection if the underlying protocol is modified, and/or decryption is required on whole or part of the encrypted traffic.
It can further be divided into two sub-categories:

\begin{itemize}
    \item {\em Partial inspection.} A scheme provides active inspection, but only in a restricted manner. For example, a few schemes based on searchable encryption technique enable only exact matching, and not regular expression type of analysis.
    \item {\em Full inspection.} A scheme provides full active inspection. This means the scheme enables all functionality as in inspection on plain data.
\end{itemize}

\end{itemize}
\bigbreak

\subsection{Man-in-the-Middle}
\label{subsec:MitM}
The {\em Man-in-the-Middle} (MitM) technique is commonly deployed in enterpise solutions for encrypted traffic analysis, for example as presented in~\cite{CarnavaletM16,DurumericMSBSBB17,WakedMY18,Broadcom20}.
There are also widely available open source tools such as MitMProxy~\cite{MitMProxy18} and SSLSplit~\cite{SSLSplit18}.
MitM uses {\em active inspection}.
The main idea is to enable the middlebox provider to act as the server endpoint so that it can decrypt, inspect and then re-encrypt network traffic of the client.
The traffic is then forwarded to the server endpoint.
In general, for a client-oriented model (Figure~\ref{fig:MBArch1}), this is achieved by first installing a certificate of the MB provider in the client device (e.g. trusted CA store for browsers).
When a client initiates a secure session with the server, the MB provider hosting the MBs intercepts the traffic and forges a certificate using the server credentials.
In this way the service provider masquerades as the server, setups a session with the client, decrypts and inspects the data.
After inspection, the service provider initiates on behalf of the client a secure session with the server.
In an enterprise network, installation of the root certificate on every client device can be performed by the network administrator.
The inspection can be performed by MBs hosted locally in the premise of enterprise network or through outsourced MB service.

\begin{figure}[!ht]
	\centering
	\includegraphics[scale=0.30]{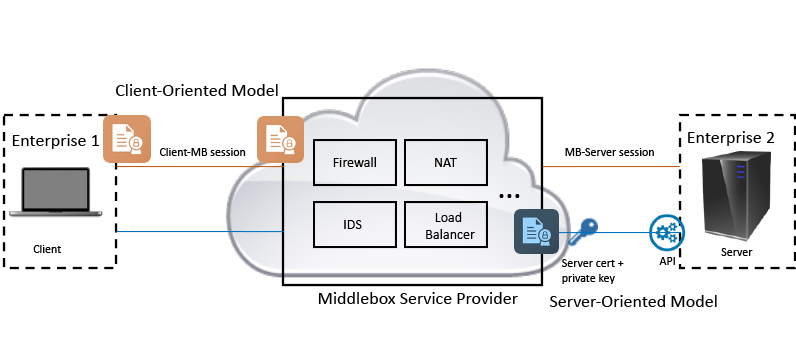}
	\caption{MitM technique: (1) Client-Oriented: Middlebox service provider installs root certificate on client devices to enable middlebox to inspect the encrypted traffic; (2) Server-Oriented in the content delivery network context: Server shares certificate and private key with the service provider, or provides a program interface for the service provider to access the private key.}\label{fig:MitM}
\end{figure}

In the server-oriented model (Figure~\ref{fig:MBArch2}), similar approach can be deployed.
Nevertheless, if the server subscribes to a content delivery network (CDN), the conventional approach is to pass the server certificate, together with the secret key to the CDN provider~\cite{BhargavanBDFO18}.
In more recent approach, the secret key is not given, but a program interface is provided to the CDN provider to have access to the key. 
There are various key sharing and content delegation approaches as discussed in~\cite{Carnavalet20}.
Durumeric {\em et\ al.}~\cite{DurumericMSBSBB17} listed some commercial solutions and explored the security issues of these solutions.
They found that nearly all the solutions they investigated have reduced connection security and five of the solutions contain severe vulnerabilities.
The technique can be coupled with a policy engine, in which a whitelist of websites can be created by an administrator so that these websites will not be inspected by the MBs~\cite{Symantec18}.
This serves as an approach to mitigate privacy concerns, especially for financial transactions such as online banking and purchases.
Nevertheless, the administrator has to be fully trusted to configure the policy engine correctly and honestly.
Figure~\ref{fig:MitM} illustrates the MitM settings for client-oriented MB model and the server-oriented MB model (for the use cases involving content delivery network). Here, we use the outsourced MB architecture since it can be used for all three system models that we have presented.
Furthermore, it is the emerging direction in both industry and research. 

In the followings we discuss the main advantages and limitations of the MitM approach.

\begin{itemize} [leftmargin=*]
\item {Advantages:}
\begin{itemize} [leftmargin=*]
    \item {\em Changes to TLS:}
    The technique can be deployed straightforwardly without changing the underlying protocol (e.g. TLS) since the service provider sits in the middle and manages secure sessions between the client-service provider and service provider-server. 
    \item {\em Functionality:}
    The MB decrypts the encrypted traffic.
    This means after the data is decrypted, it is able to run similar functions as in the case where the traffic is not encrypted.
    Hence it maintains full functionality as if the traffic is not encrypted.
    \item {\em Performance:}
     It is relatively efficient compared to the other techniques that require additional cryptographic computations, machine learning operations or computation in the trusted hardware.
     As most modern CPUs support AES-NI instructions, the time to encrypt or decrypt a block of bits take 3$\mu$s per packet of 1500 bytes. According to \cite{SherryLPR15}, the vanilla TLS setup phase takes 73ms on a 20 Mbps throughput network. 
     Certainly, it incurs higher overhead compared to inspection on plain data since the MB is required to decrypt and re-encrypt the traffic.
     For commercial solutions, the middlebox latency can be less than 40$\mu$s, and depending on the device variant, TLS inspection throughput ranges from 250Mbps-10Gbps.
     These are based on publicly available information from the enterprises that we surveyed~\cite{ssl-visibility,gigamon-visibility}. 
     \end{itemize}
\item{Limitations:}
    \begin{itemize} [leftmargin=*]
    \item {\em Security:}
    The main issue in MitM is that the MB provider learns the content of the traffic and this may cause privacy concern, for example, for an enterprise that subscribes to an outsourced MB service.
    Another issue is that we must trust the MB providers to implement the underlying protocol correctly and uses the up-to-date version of the protocol. 
    As discussed in Section~\ref{sec:Introduction} and Section~\ref{subsec:IndustryPractices}, weaknesses were found in some of the solutions due to implementation of the TLS protocol using old versions or deprecated ciphers.
    In addition to this the gateway (or the service provider) becomes highly valuable target for attacker and the administrator must be fully trusted.
    \end{itemize}
\end{itemize}



\subsection{Searchable Encryption}
\label{subsec:SE}
\begin{figure}[!ht]
	\centering
	\includegraphics[scale=0.30]{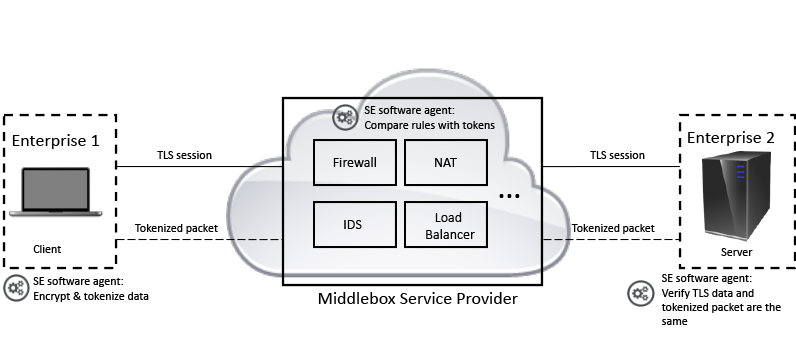}
	\caption{SE technique: All models have the same setting. The main characteristic is that there are two data flows, one the TLS session and another the tokenized data. MB inspects the tokenized data only, and forward the TLS session to the server. The server is able to verify the two data flows are identical since it has the session key used to encrypt the TLS traffic and the tokenized data.}\label{fig:SE}
\end{figure}

A second technique that was proposed is to detect malicious traffic via token matching without decrypting the underlying encrypted traffic.
We categorise the technique as {\em searchable encryption} (SE) because the main idea is to use searchable encryption scheme to map between encrypted keywords and the encrypted rulesets.
SE uses {\em active inspection}.
{\bf Sherry {\em et\ al.}~\cite{SherryLPR15}} introduced {\bf BlindBox}, the first privacy-preserving deep packet inspection scheme using SE technique.
In BlindBox, a client initiates a TLS session with the server, as well as another connection for token matching.
Both the connections route through the MB.
The general idea is that the MB hosts rulesets that are encrypted using a key derived from the session key of the TLS session.
The client then tokenizes and encrypts its message using the same key, and transmits the tokenized traffic through the second connection.
The MB then try to match the tokenized traffic with the encrypted rulesets.
If there is a match, the traffic is considered malicious and blocked.
Figure~\ref{fig:SE} shows a general setting for SE-based schemes.

The MB should not know the key used to encrypt the rulesets and to generate the token since if this is the case, it is able to decrypt all tokens in the tokenized traffic and learns the message.
This is akin to having access to the plaintext.
BlindBox mitigates this issue by performing encryption of the rulesets through garbled circuit and oblivious transfer for every session.
This means BlindBox is computationally expensive, at least in the initial handshake of the TLS protocol.
Furthermore, the scheme must be able to verify that the tokenized traffic is identical to the TLS session. This is because the client can send malicious message through the tokenized traffic and a benign one through the TLS session, thus defeating the inspection.
Here, verification is performed by the server.
The server receives data from both connections, and since the server has the session key, it can decrypt the TLS traffic as well as generate the token to verify that the content from both connections are the same.
However, inspection based on token matching of the encrypted traffic is rather limited as it would not be able to cater for regular expression inspection.
BlindBox proposed a probable cause privacy mechanism to mitigate this issue by allowing decryption of the underlying traffic based on the session key (without revealing the session key). 

The BlindBox mechanism is extended to the setting of outsourced MB by {\bf Lan {\em et\ al.}~\cite{LanSPRL16}}.
The scheme is termed {\bf Embark}.
It enhanced the token matching technique to include prefix matching, as well as different searchable encryption functionalities catering specifically for different MB services such as IP firewall, NAT, HTTP Proxy, data exfiltration and intrusion detection.
It also proposed a different, but practical setting.
In order to mitigate the computation overhead of BlindBox, {\bf Canard {\em et\ al.}~\cite{CanardDKPS17}} proposed {\bf BlindIDS}.
It uses a different approach in that pairing-based public key operation is deployed for token matching and for the client to communicate with the server through the MB.
This also means existing TLS protocol must be replaced with their scheme.
Another scheme that uses public key operation is a scheme termed {\bf SPABox} by {\bf Fan {\em et\ al.}~\cite{FanGRCQ17}}.
In contrast to BlindIDS, the public key operation based on homomorphic encryption is used only to perform machine learning based inspection.
The scheme also uses Diffie-Hellman based oblivious pseudo-random function for encrypted rule preparation, which is performed interactively between the MB and the server.
For regular expression matching, the scheme deploys a variant of garbled circuit that is more efficient than the general construction.
{\bf Yuan {\em et\ al.}~\cite{YuanWLW16}} proposed a scheme that is more efficient than BlindBox using a high-performance encrypted filter.
Their scheme also extends the token matching mechanism of BlindBox to work for multi-condition rulesets.
Though efficient, their scheme requires the server to first register with the administration service of the enterprise hosting the client.
Only then the client may initiate the TLS session with the server.
The performance of BlindBox is improved significantly by {\bf Ning {\em et\ al.}~\cite{NingPLCC19}} in their scheme termed as {\bf PrivDPI}.
The computation time during the initial handshake was greatly reduced when compared to BlindBox, and a reusable obfuscation mechanism generates intermediate values that can be reused across subsequent sessions for repeated tokens, which could further speedup token encryption.
The efficiency of PrivDPI is further improved in a scheme called {\bf Pine}, which is also proposed by {\bf Ning {\em et\ al.}~\cite{NingHPXLWD20}}. In addition, Pine enables rule hiding as middlebox services migrate to third-party cloud setting, rule privacy may be a concern.
{\bf Ren {\em et\ al.}} proposed {\bf EV-DPI~\cite{ren2020privacy}}. EV-DPI is a two-layered architecture design and is deployed over two non-colluding servers to outsource the middlebox. First layer filters out legitimate packets using encoded bloom filters. The second layer supports exact rule matching for packet inspection using conjunctive searchable encryption scheme proposed by lai {\em et\ al.}~\cite{lai2018result}. The scheme allows inspection results to be efficiently verified using cuckoo hashing.
A related scheme that uses a different approach under the cloud setting for NFV is SplitBox proposed by Asghar {\em et\ al.}~\cite{AsgharMSCKM16}.
They suggest the use of two cloud systems where every rule in the rulesets is XOR with a random string and split into many blocks to the various MBs resided in one of the cloud system.
Both cloud systems collaboratively compute the blocks in order to perform inspection on the traffic.

\begin{itemize} [leftmargin=*]
\item{Advantages:}
\begin{itemize} [leftmargin=*]
    \item {\em Changes to TLS:}
    The main advantage of this technique is that modification on the underlying protocol (i.e. SSL/TLS) is not required (except for BlindIDS which proposes a new protocol that may replace SSL/TLS), yet it allows privacy-preserving inspection.
    Some of the constructions can also be efficient in specific scenarios.
    As an example, assuming a trusted rule generator (which is the case for a few of the schemes stated in Table~\ref{tbl:YES}), encryption of the rulesets can be performed efficiently if it can be assumed that the rulesets are encrypted by the client or another trusted third party and passed to the MB.
    Inspections require only comparing the rulesets with the tokenized data, and does not require decryption and re-encryption as in the MitM approach.
    \item {\em Security:}
    It provides better security guarantee than the MitM technique and the access control technique (Section~\ref{subsec:AC}) since the inspection is performed without decrypting the encrypted traffic.
    Hence the MB and the service provider would not be able to learn the content of the traffic, except for those tokens that match the rules. An exception is that there are schemes such as BlindBox, PrivDPI and EV-DPI which follows probable cause privacy where the middlebox gets to see the decrypted traffic if and only if the flow is deemed to be suspicious. 
\end{itemize}
\item{Limitations:}
    \begin{itemize} [leftmargin=*]
    \item {\em Functionality:}
    The main drawback is that utility can be limited due to simple mapping of encrypted tokens and rules.
    This may not be sufficient especially for inspection requiring regular expression.
    Some of the schemes address this problem by allowing the MB to decrypt the traffic when there is a match.
    Alternatively, compute-intensive primitive such as homomorphic encryption is used to enable expressive inspection.
    Furthermore, in may cases a separate channel is required to communicate the tokenized traffic, as shown in Figure~\ref{fig:SE}.
    For the MB to be able to inspect inbound and outbound traffic, both endpoints, meaning the client and the server must have the scheme installed.
    This is in contrast to the MitM technique (as well as the machine learning and trusted hardware approach, which we will discuss in the subsequent sections). 
    \item {\em Performance:}
    At the client, there is the compute-intensive overhead due to the additional operations of tokenizing and encrypting the data stream compared to pure TLS connection, and the MitM technique.
    At the MB, however, only matching is performed. 
    Hence, generally the setup phase for SE techniques contribute more to the overall performance delay than the inspection phase. For BlindBox, the setup time is 97s for 3000 rules and 33$\mu$s for inspecting 1 packet with 3000 rules. BlindIDS limits the setup time to 73ms but increases the inspection time to 74s for same setting. By replacing garbled circuits with reusable exponentiations in group, PrivDPI prepares 3000 encrypted rules in 0.64s which is 288 times more efficient than BlindBox. PrivDPI also reduces the bandwidth usage(from GBs to KB) by replacing the garbled tables with reusable obfuscated rules. Pine further improves the one-round connection time of PrivDPI with hash operations. For SPABox, lack of hardware support for operations results in slower(9 times than BlindBox) performance for encrypting a token at setup phase. However, SPABox saves on average 29.5\% time in inspection phase compared to BlindBox due to hashing technique. 

    \end{itemize}
\end{itemize}

%

\subsection{Access Control}
\label{subsec:AC}
\begin{figure}[!ht]
	\centering
	\includegraphics[scale=0.30]{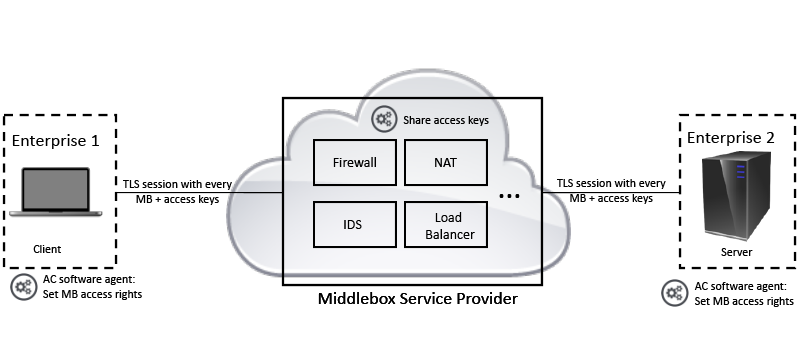}
	\caption{AC technique: All models have the same setting. The main property is that MBs are visible to the endpoints and that the endpoints may decide access rights of the MBs to the encrypted traffic. The access rights are controlled using read/write keys}\label{fig:AC}
\end{figure}

This is a technique that advocates client and server accountability, where the client and the server are aware of all the MBs deployed in between the two of them.
Also, they are given the ability to authenticate and assign access rights to these MBs.
Hence the term {\em access control} (AC).
AC uses {\em active inspection}.
Figure~\ref{fig:AC} illustrates a typical setting for AC-based schemes.

{\bf Naylor {\em et\ al.}~\cite{NaylorSVLBLPRS15}} introduced a scheme of this type, termed {\bf mcTLS}. 
It modifies existing TLS protocol to allow the client, the MBs and the server to establish
secure and authenticated channel, and exchange read and write secret keys in addition to the session key.
The mcTLS protocol partitioned the network traffic according to a least privilege MB access policy.
Each MB is assigned a read and/or a write key.
This involves providing different level of access control to read and/or write on the encrypted traffic.
The encrypted traffic is partition so that some portions of the payloads can be decrypted.
The benefit of such a technique is that there is a fine-grained control and flexibility in preserving privacy on portions of the data that should not be decrypted.
In other words, mcTLS only partially preserve privacy of the underlying payload, since a MB has the keys to decrypt the portions of the traffic where authorisation is given.

The main issue with mcTLS is that it is a new protocol.
For adoptions, existing TLS protocol that is widely used must all be replaced with mcTLS.
Acknowledging this issue, {\bf Naylor {\em et\ al.}~\cite{NaylorLGKS17}} further proposed another scheme, termed {\bf mbTLS}.
It does not change the underlying TLS protocol, except in introducing extensions that can be readily adopted using existing protocol.
It is designed to be scalable for implementation in the outsourced MB setting.
In more details, mbTLS considers the problem of outsourcing MB functions to multiple third parties. For example, there maybe multiple MBs contracted out by a client, in addition there can also exist a set of MBs for the server, all residing in untrusted cloud environment. They discussed several important properties desired in such a scenario, and designed mbTLS to meet some of such important properties. In mbTLS, a different TLS session exists between every two entity (client, MB, server), with each session operating using its own secret key. This design is such that, the client is only aware of the client-side MBs and none on the server side; and vice versa. The MBs in the cloud uses Intel SGX to protect the data and keys from the cloud infrastructure provider. The initial control messages between the different entities are multiplexed over a single TCP connection, thus adding no additional round trip time.  
In mbTLS, MBs are able to view the entire payload, as the sessions are handed over to participating entities.  

{\bf Bhargavan {\em et\ al.}~\cite{BhargavanBDFO18}} demonstrated attacks on mcTLS, and proposed a formal model on analyzing the protocol.
In brief, one of the attacks exploit the fact that mcTLS does not perform authentication and the handshake finished message for the session initiation between the client and the MB, and between the MB and the server.
Due to this, Bhargavan {\em et\ al.} further proposed a protocol that addresses these attacks under the security model that they defined.
It does not require any changes to the TLS protocol, in contrast to mcTLS.
In response to the attack, the ETSI draft standard that is based on mcTLS proposed a fix to the mcTLS based on message authentication code in their {\bf MB Security Protocol (MSP) specification}~\cite{ETSIDraftII18}.
At the moment, it remains to be seen how the industries will accept a secure but MB-friendly new protocol for encrypted traffic, since it may mean replacing TLS with MSP.
{\bf Lee {\em et\ al.}~\cite{LeeSLCCCK19}} proposed {\bf maTLS}, a middlebox-aware protocol to address the MitM pitfalls. Middleboxes are made visible so that server can be explicitly authenticated, the encryption parameters used can be verified and whether messages are modified. Every middlebox is issued a certificate by a CA, and the certificate is logged in a  middlebox transparency log server. Through the log server, a middlebox certificate is publicly verifiable and revocable. However, as stated in~\cite{Carnavalet20}, while the middlebox certificates are vetted by the CA, the client (or end user) still need to decide to accept these certificates. An attacker may launch a phishing attack by obtaining a certificate with convincing name. 


\begin{itemize} [leftmargin=*]
\item{Advantages:}
\begin{itemize} [leftmargin=*]
    \item {\em Security:}
    The main provision of the AC technique is that it provides {\em accountability}.
    It means the endpoints will be able to authenticate the deployed MBs, which is unique to this technique.
    
    \item {\em Performance:}
    The technique does not require special cryptographic primitives as in SE and hence is more efficient.
    It is less efficient as the MitM approach, but provide more flexibility in terms of preserving privacy of the data.
    It does not require installation of certificate at the client, nor allowing any of the MBs to decrypt and view the full payload of the traffic.
    For example in mcTLS, the performance during handshake protocol depends on number of contexts and middleboxes. Time required for the first byte to reach the endpoint is 400ms with 10 contexts and 560ms with 14 contexts. 
    
    \item {\em Functionality:}
    It provides full functionality as in the MitM technique.
    This is because for a specific MB, such as an IDS, access would be given to decrypt the portions of the encrypted traffic for the IDS to perform the required inspections.
    \end{itemize} 
 \item{Limitations:}  
    \begin{itemize} [leftmargin=*]
    
    \item {\em Changes to TLS:}
    The main pitfall of this approach is it would require all parties, meaning the two endpoints and all the MBs to fully co-operate in order for the scheme to work.
    This is because the client, and the server must know the type of MBs that they are communicating with.
    The client, the server and all the MBs must agree on the scheme to be used.
    It is also not clear how the client and the server may define context (which portion of the data can be reveal, to which MB), especially for data in different domain and then set the access policy for these MBs.
    
    \item {\em Security:}
     Union of the access read/write keys for the scenario whereby MBs are hosted at the service provider will allow the service provider to decrypt most probably all portions of the encrypted traffic.
    \end{itemize}
\end{itemize}

%
%
%

\subsection{Machine Learning}
\label{subsec:ML}
\begin{figure}[!ht]
	\centering
	\includegraphics[scale=0.30]{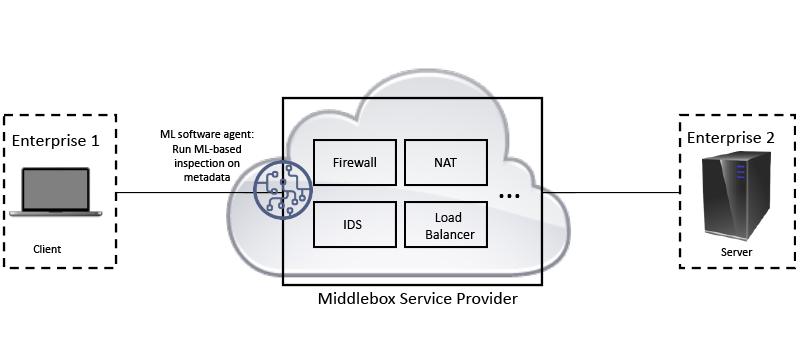}
	\caption{ML technique: All models have the same setting. The main property is that no changes are required to existing industrial deployment on the endpoints. Inspections are performed without needing to modify or decrypt the encrypted traffic.}\label{fig:ML}
\end{figure}

Inspecting encrypted traffic based on machine learning (ML) technique represents an ideal solution in terms of security and application setting since it does not require any changes to the existing setup.
The idea is to analyse the plain metadata and header of the protocol, extract features from the encrypted payloads, as well as analyse telemetry data from the network traffic.
For examples, throgh observing behavioural properties (e.g. the round trip time, number of packets sent), observing the encrypted payloads, and observing additional information such as protocol handshakes. 
It has been shown to be effective in particular for use cases such as traffic clustering, application type and protocol classification, anomaly detection and file identification~\cite{ETAENISA19}.
Among all the techniques, it is the only technique that uses {\em passive inspection}.
Figure~\ref{fig:ML} shows a high-level setting of the ML technique.

There are many proposals using ML technique for use cases such as for application type and protocol classification, anomaly detection and file identification. These were comprehensively surveyed in~\cite{ETAENISA19}.
Another proposal for anomaly detection is the technique proposed by {\bf Yamada {\em et\ al.}~\cite{YamadaMTSP07}}.
Their scheme performs anomaly detection using only size of data and timing information of the encrypted traffic.
More recently, {\bf Anderson {\em et\ al.}~\cite{AndersonM16,AndersonM17,AndersonPM18}} proposed techniques for malware detection that in addition of utilising sequence of data size and timing, also uses the various TLS header information and DNS data.
Anderson {\em et\ al.} demonstrates that some of the malware-based encrypted traffic possess distinct characteristics compared to enterprise network traffic.
These characteristics when combined with other telemetry data, allows accurate classification of the malware traffic.
They have also showed that random forest method outperforms other methods in terms of classifying malware traffic~\cite{AndersonM17}.
Yet by careful engineering of the features, including recommendations by domain experts, linear regression using more expressive features actually outperforms random forest method.

There are also techniques that identify malware traffic based on fingerprinting.
For example, {\bf JA3/JA3S} proposed by {\bf Althouse {\em et\ al.}~\cite{JA320}} generates malware fingerprints based on the TLS metadata (e.g. handshake messages). {\bf Anton~\cite{Anton20}} proposed a technique that creates rules by observing the packet byte stream of the traffic to create fingerprint. 
Specific malware can be detected based on the unique packet byte stream transmitted from the client to the server.
The technique is specifically created for use with Suricata.
Both the above proposals can be considered as signature-based techniques and require prior knowledge of the malware.



\begin{itemize} [leftmargin=*]
\item{Advantages:}
\begin{itemize} [leftmargin=*]
    \item {\em Changes to TLS:}
    The main advantage of the ML technique compared to all the other techniques is that it does not require any changes to existing encrypted traffic setup.
    No modification is required at the client or the server.
    The MB installs the ML module so that it is able to analyze the metadata of the encrypted traffic.
    
    \item {\em Security:}
    Since no changes are required to the setting and the underlying protocol, it preserves the security guarantee of the original setting.
    For example, using ML technique maintains end-to-end encryption of a TLS session between the client and the server.
    This represents another main advantage compared to all other techniques that require changes to the protocol (i.e. AC technique) or changes to the client and server settings (i.e. MitM, SE, and trusted hardware).
    However, we remark that machine learning-based analysis and fingerprint techniques may also be used to learn information about a user.
    For example, it is possible for an attacker to learn about the websites a user surfs or find out the files a user downloads and shares even though the traffic is encrypted~\cite{ETAENISA19}.
    It is an interesting area of study as to how pervasive ML techniques can also be used to learn about an entity, in contrast to the techniques being used as a privacy-preserving approach for analysing encrypted traffic without knowing the payloads.
 \end{itemize}   
 \item{Limitations:}
    \begin{itemize} [leftmargin=*]
    \item {\em Functionality:}
    The main concern is whether the technique is sufficiently comprehensive to cover most of the detection requirements of MBs performing security functions.
    As was stated in~\cite{ETAENISA19}, there are inherent limitation to what can be analysed based on Ml techniques, and as was discussed in~\cite{Carnavalet20}, there are use cases that will require inspection on the payloads, not just the headers and metadata.
    As of now, the most successful ML mechanism of Anderson {\em et\ al.}~\cite{AndersonM16,AndersonM17,AndersonPM18} only cater for malware detection.
    All in all, the technique will need to be able to cater for different types of middlebox services.
    
    \item {\em Performance:}
    It remains to be seen how the performance of the technique as compared to other techniques.
    Training must be performed before detection can be carried out on real-time data.
    The timely dataset and accuracy of the training model are area to be further explored.
    As was discussed by Anderson and McGrew~\cite{AndersonM17} in their work on malware classification using machine learning technique, it took approximately $200$ seconds to train and less than $10$ seconds to test.
    This is for their best performed random forest algorithm, on the enhanced feature.
    The timing is a rough estimation from Figure 5 in their work~\cite{AndersonM17}.
    Also, solutions based on this approach need to continuously feeding high fidelity labelled data for training, which can be hard to obtain.
    
    \end{itemize}
\end{itemize}

%
  %
 %

\subsection{Trusted Hardware}
\label{subsec:TH}
\begin{figure}[!ht]
	\centering
	\includegraphics[scale=0.30]{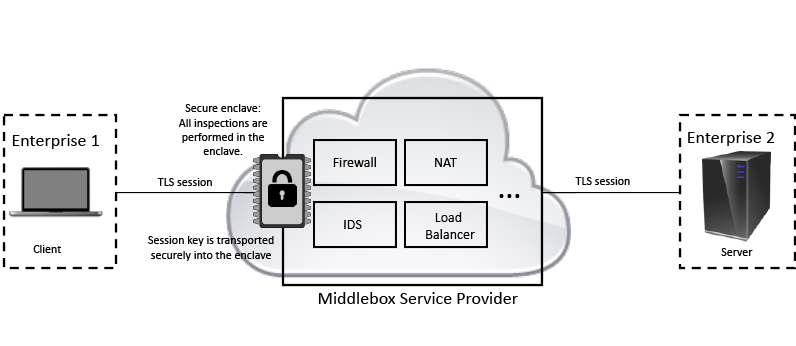}
	\caption{TH technique: Similarly all models have the same setting. No changes are required on the TLS protocol. The main property is that decryption of the traffic is performed in the secure enclave. The main requirement is on securely transporting the session key into the enclave, either by the client or the server.}\label{fig:TH}
\end{figure}

Trusted hardware (TH) has also been deployed for privacy-preserving packet inspection.
The emerging practice is to utilize the secure enclave of the Intel SGX trusted hardware.
The general idea is for the client or the server to share the session key securely with the enclave residing in the MB.
The decryption, inspection and re-encryption is performed in the enclave.
The MBs and the service provider hosting these MBs are not able to discern or learn the data and processes.
The challenge is how to implement middlebox functionalities efficiently and securely to fully utilize the capability of the trusted hardware.
We may classify this technique as using {\em active inspection}, but the inspection is hidden from the view of the entity hosting the trusted hardware.
Figure~\ref{fig:TH} shows a high-level setup of the TH technique.

{\bf Han {\em et\ al.}~\cite{HanKHH17}} proposed a scheme, {\bf SGX-Box}, using this technique.
In their scheme, the server shares the session key with the Intel SGX's secure enclave through an out-of-band secure channel.
They presented an efficient implementation of performing the inspection inside the secure enclave.

It is also possible to push the MB functions to the network edge (or to the client's device).
This is anologous to hosting the MB at the client, except that inspection are performed in the secure enclave embedded in the client device.
This is beneficial in that all outbound traffic is analyzed before it is transmitted from the client device.
Inbound traffic on the other hand is verified before it is released to the client device.
{\bf EndBox}, proposed by {\bf Goltzsche {\em et\ al.}~\cite{GoltzscheRN+18}}, provides such a solution.

There are also works that propose comprehensive secure Network Function Virtualization (NFV) systems using TH technqiue.
These NFV systems provide inspection of encrypted traffic as one of their functionality.
For example, {\bf SafeBricks} proposed by {\bf Poddar {\em et\ al.}}~\cite{PoddarLPR18}. 
SafeBricks provides MB as a cloud service construction.
It ensures the cloud provider only sees the encrypted traffic, where the application header (normally visible under TLS) is also encrypted based on IPSec.
Additionally, it shields the rulesets and the network function codes from the cloud provider.
LightBox further improves on existing TH-based schemes, including SafeBricks, in terms of efficiency and properties.
For instance, LightBox additionally protects privacy of metadata of the traffic, including packet size, count and timing. 
Two related works are ShieldBox proposed by Trach {\em et\ al.}~\cite{TrachKGABF18} and LightBox proposed by Duan {\em et\ al.}~\cite{DuanYW17}. ShieldBox and LightBox provide secure middlebox functionalities by provision of NFV systems in SGX. 
However, both are geared towards efficient deployment of MBs on untrusted cloud providers and would require adaptation to specifically cater for encrypted traffic inspection.

\begin{itemize} [leftmargin=*]
\item{Advantages:}
\begin{itemize} [leftmargin=*]
    \item {\em Changes to TLS:}
    One advantage is that it does not require changes to the underlying protocol.
    However, it must be able to transport the session key to the trusted hardware securely, either by the client or the server.
    
    \item{\em Functionality:}
    Another advantage is it provides MitM-type full functional inspection without the drawbacks of the MitM approach.
    It allows a MB to inspect the encrypted traffic in a shielded environment, so that no decrypted packets are leaked.
    It may represent the most practical solution if hardware deployment, cost and security of the trusted hardware is not an issue.
    \end{itemize}
\item{Limitations:}
    \begin{itemize} [leftmargin=*]
    \item {\em Security:}
    The current main concern is that security of the trusted hardware (i.e. Intel SGX) is still being actively study.
    As was presented by Lindell~\cite{Lindell18}, there has been successful side-chanel attacks on Intel SGX.
    It remains to be seen how serious such attacks are in practical deployment.
    
    \item{\em Performance:}
    It is less efficient for inspection, at least when compared to SE scheme as discussed in~\cite{LanSPRL16}.
    Take SafeBricks for example, it introduces 16\% bandwidth overhead when compared to Embark's 21\% bandwidth overhead. Nevertheless, SafeBricks also incurs 0-15\% throughput performance overhead at the middlebox due to SGX compared to negligible overheads of Embark and BlindBox. However, Embark(at gateway) and BlindBox(at client) incur high latency in order to encrypt the packets at the endpoints, where SafeBricks does not pay such cost.
    Another issue is the availability of the hardware and cost.
    While most recent devices would have a secure enclave embedded (i.e. Intel SGX), issues remain for those that are not, especially legacy systems.
    \end{itemize}
\end{itemize}

\subsection{Comparisons}
\label{subsec:compare}
As a summary for what we have discussed thus far, Table~\ref{tbl:compare} provides comparisons between the models, use cases, characteristics (e.g. security, types, models, outsourced MB to cloud), the five techniques and performances.

\subsubsection{Security}
In the following we define three different levels of security, shown in Table~\ref{tbl:compare}.  The main goal of the schemes is to provide data privacy (as stated in Section~\ref{subsec:secReq}).
In other words, how much information is leaked to an adversary, which may include the entity hosting the MBs.
 
\begin{itemize} [leftmargin=*]
	
	\item {\em Full Reveal} ($\Circle$).
	This means the network traffic or data payload is visible (or decrypted in total) to the MB.
	Solutions based on MitM technique are in this category.
	
	\item {\em Partial Reveal} (\pie{180}, \pie{270}). 
	This means only partial content of the network traffic is revealed (or partial decryption of the payload) to the MB.
	Solutions based on SE and AC techniques are in this category.
	However, there is a subtle difference between the partial data exposure between the two techniques.
	SE matches encrypted rules with the encrypted token generated from the payload, and hence only reveal matched result.
	Encrypted payload that does not match any of the rules remain private from the view of the MB.
	An exception are schemes that reveal the underlying data for more complex inspection such as regular expression.
	In contrast, schemes based on AC technique directly divulge the underlying payload as long as the MB has the access right to some parts of the encrypted traffic data.
	By the above observation, we may state that SE techniques leak less information (\pie{270}) when compare to AC techniques (\pie{180}).
	
	\item {\em Hidden} ($\CIRCLE$).
	This means the network traffic or data payload is hidden (or remain encrypted) and the MB has no visibility to the underlying content.
	Schemes based on ML and TH techniques are in this category.
	Note that here we assume the trusted hardware deployed in the schemes are considered secure, in that there is no information leakage from the hardware itself when executing the schemes.
	This may be a strong assumption as there are on-going work demonstrating side-channel attacks to such hardware implementation~\cite{Lindell18}.
	It remains a research problem as to the effects of such attacks toward schemes using TH technique.  
	
\end{itemize}

\subsubsection{Utility}
In terms of utility, we define two broad categories, which are also shown and described in Table~\ref{tbl:compare}. 

\begin{itemize} [leftmargin=*]
	\item {\em Full functionality ($\CIRCLE$).}
	It means a scheme provides inspection utilities similar to that of inspection on plaintext traffic.
	MitM approaches are in this category.
	
	\item {\em Partial functionality (\pie{90}, \pie{180}, \pie{270}).}
	It means a scheme only provide partial utilities. For example, solutions based on SE perform direct matching and prefix matching against pre-defined rule sets. They also require specific setting such as tokenisation of the messages and a separate encryption channel in addition to the TLS traffic. Hence functionality for SE techniques is limited when compared to other techniques (\pie{90}). 
	ML-based solutions also have limited inspection capability as was previously discussed~\ref{subsec:ML}, but they do not require any changes to the existing setting and can be deployed directly (\pie{180}).
	Solutions based on AC and TH in theory can achieve full inspection utilities as in MitM approach. The AC context and policy can be set to authorise a middlebox to have full control on the traffic (i.e. decrypt then read and write on the messages), and the TH in principle can decrypt and perform all types of inspection under the secure enclave. 
	However, in practical terms, the main goal of the AC technique is to restrict access to the encrypted payload depending on the requirement of a middlebox. By doing so the client and the server may control the portion of the data that is allowed to be read or modify by a middlebox. In contrast to the MitM approach, the setup is such that no one middlebox is able to decrypt the full message, except for the recipient. It would also require new setting and configuration of access context for the client and the server. 
	For TH, the secure enclave provides limited storage/memory and computation capabilities.
	It would not be practical to perform all types of inspection under the secure enclave, at least not for the current TH technology. 
	Hence we denote both the AC and TH techniques as achieving partial but better functionality (\pie{270}), as compared to the SE and ML-based solutions.
\end{itemize}

%
%
\section{Discussions}
\label{sec:ChallengesNResearchDirections}
For all the novel schemes that have been proposed to date, a question that one may ask is why is that most of the industries still prefer and deploy MitM-based solutions.
Conversely, if a MitM-based solution is designed in a careful manner, that is, with up-to-date TLS configuration and policy engine, is it not sufficiently secure?
On the other hand, schemes based on the four techniques were and are continuously being proposed to address the issues with the MitM approach, i. e. violation of end-to-end encryption, privacy concern and difficulties or flaws in the configuration and settings.
The main goals are mainly to maintain end-to-end security provided by the underlying protocol such as TLS, and reduce the information revealed to the MB.
One of the reasons is to reduce the trust placed on the MB providers so that one does not rely on the provider to {\em design and implement encrypted traffic inspection in a careful manner}, which is always not possible as demonstrated by Jarmoc~\cite{Jarmoc12}, Carnavalet and Mannan~\cite{CarnavaletM16}, and Durumeric {\em et\ al.}~\cite{DurumericMSBSBB17}.
{\em All in all the aim is to replace the MitM-based solution with either of the other four techniques (or combination of them) for better privacy preservation yet maintain similar utility and practicalility in deployment.}

We discuss the challenges in the techniques discussed and suggest research directions.
Our discussions are based on common properties in the cyber security landscape, that is, security, performance and utility.

\subsection{Security}
	
\paragraph{\bf Information Leakage.}
The main challenge in terms of security is leakage of information in the techniques that we have discussed.
In schemes based on SE, token matching may still leak information if the client or the service provider (in the case that the provider should not know the rulesets) has background information of the underlying rulesets.
This is a reasonable assumption since there are publicly available rulesets such as the rulesets provided by Snort.
This concern was discussed by Poddar {\em et\ al.}~\cite{PoddarLPR18}.
Furthermore, there are well-established attacks on searchable encryption schemes such as inference attack~\cite{IslamKK12}, leakage-abuse attack~\cite{CashGPR15}, reconstruction attacks~\cite{KellarisKNO16} and passive attacks~\cite{NingXLZC19}.
Similarly, further study is required on schemes that utilise trusted hardware due to the recent attacks based on side information on the secure enclave, such as the attacks discussed in~\cite{BulckMWGKPSWYS18}.
	
\paragraph{\bf Potential collusion in the outsourced MB scenario.}
In terms of the outsourced MB model, one of the challenges that has not been examined is the possibility of collusion between a malicious client or server and the cloud service provider.
The malicious client or server may collude with the service provider by providing the session key in order for the provider to decrypt the underlying traffic, thus circumventing the executions of the schemes, be it through  solutions based on SE, AC, ML or TH.
A new security model may be required to model such scenario and the question would then be whether this is a practical assumption.
For example, in the scenario where an enterprise outsources its content management to a content distribution network (CDN) provider, the enterprise shares the certificate private key, or delegates content to the CDN provider.
Assuming that the CDN provider uses its own cloud infrastructure, there is no collusion in this instance.
In contrast, if an enterprise outsources network threats detection to a managed service provider, whereby the provider uses a third-party cloud provider to host its various MB services, then the possibility of collusion might need to be taken into consideration.

\paragraph{\bf Challenges for ML Technique.}
ML technique enables analysis without needing to change and decrypt the existing encrypted network setting (e.g. TLS).
This provides the best solution since it passively inspect traffic.
Nevertheless, the one challenge is whether the technique is sufficiently comprehensive to cover most of the detection requirements of MBs performing security functions.
As of now, the most successful ML mechanisms of Anderson {\em et\ al.}~\cite{AndersonM16,AndersonM17,AndersonPM18} only cater for malware detection.
Also, training must be performed before detection can be carried out on real-time data.
The timely dataset and accuracy of the training model are area to be further explored.

\afterpage{%
    \clearpage
    \thispagestyle{empty}
    \begin{landscape}%
    
\begin{table*}
	\centering
	\begin{adjustwidth}{-5cm}{}
	\caption{Privacy-Preserving Techniques for Encrypted Traffic Inspection: Types, Techniques and Applications \label{tbl:compare}}{
		\scriptsize
		\renewcommand{\tabcolsep}{0.45pc} 
		\renewcommand{\arraystretch}{1.1} 
		\begin{threeparttable}
			\begin{tabular}{@{}|l|c|c|c|c|c|c|c|c|c|c|c|l|l|l|l|}
				\hline
				Scheme 					& chg. TLS & \multicolumn{2}{c|}{Types} &  \multicolumn{4}{c|}{Technique} & Cl. & L. & Util. & Pri. & \multicolumn{4}{c|}{Configuration} \\\hline
				& & Pa. & Act. & AC & SE & ML & TH & & & & & Initial Setup & Pre-processing & Encrypt$/$Decrypt & Match$/$Inspect\\\hline\hline
				MitM approach		& $\times$ &  & $\bullet$ & & & &  &  & $\Circle$ & $\CIRCLE$ & cert. $+$ policy & Client install root cert & client$/$server share session key & MBs decrypt & MBs inspect decrypted\\
				& &  &  & & & &  &  & &  &  & Server share private key & Server delegate content & messages & payload\\\hline
				mbTLS~\cite{NaylorSVLBLPRS15}		& $\times$ &  & $\bullet$ & $\bullet$ & & & $\bullet$ & $\bullet$ & $\pie{180}^*$ & $\pie{270}$ & sym. & Client $\&$ server & Delegate MBs'  & MBs partial decrypt & MBs inspect decrypted \\
				&  &  &  &  & & &  &  &  &  &  & agrees on list of MBs & read$/$write keys & by read$/$write access & payload \\\hline
				mcTLS~\cite{NaylorLGKS17}			&  $\checkmark$ &  & $\bullet$ & $\bullet$ & & &   & & $\pie{180}^*$ &  $\pie{270}$ & sym. & Same as mbTLS$^@$ & $--$ & $--$ & $--$\\\hline
				MSP~\cite{ETSIDraftII18}			&  $\checkmark$ &  & $\bullet$ & $\bullet$ & & & & & $\pie{180}^*$ & $\pie{270}$ & sym. & Same as mcTLS$^+$ & $--$ & $--$ & $--$\\\hline
				Bhargavan {\em et\ al.}~\cite{BhargavanBDFO18}		&  $\times$ &  & $\bullet$ & $\bullet$ & & & & & $\pie{180}^*$ & $\pie{270}$ & sym. & Same as mcTLS$^\#$ & $--$ & $--$ & $--$ \\\hline
				maTLS~\cite{LeeSLCCCK19}		&  $\times$ &  & $\bullet$ & $\bullet$ & & & & & $\Circle$ & $\CIRCLE$ & sym. & MBs get certs from CA & MBs share ciphersuites & MBs decrypt & MBs inspect decrypted \\
				&  &  &  &  & & & & &  &  &  & (Certs publicly verifiable) & setting with client & messages & payload \\\hline
				BlindBox~\cite{SherryLPR15}				&  $\times$ &  & $\bullet$ & & $\bullet$ & & & & $\pie{270}$ & $\pie{90}$ & sym. $+$ SMC & Rule generator prepares $\&$  & Client tokenises messages & Client encrypts tokens & Exact match: \\
				&  &  &  & &  & & & &  &  &  &  signs rules & Client $\&$ MB jointly encrypt &   & encrypted tokens vs rules\\
				&  &  &  & &  & & & &  &  &  &   & rules &  & \\\hline
				Embark~\cite{LanSPRL16} 					&  $\times$ &  & $\bullet$ & & $\bullet$ & & & $\bullet$ & $\pie{270}$ & $\pie{90}$ & sym.  & Enterprise gateway & Gateway tokenises & Gateway encrypts & Exact $\&$ prefix match: \\
				&   &  &  & &  & & &  &  &  &  & encrypts rules, passes & messages from client & tokens & encrypted tokens vs rules \\
				&   &  &  & &  & & &  &  &  &  & encrypted rules to MBs &  &  &  \\\hline
				Yuan {\em et\ al.}~\cite{YuanWLW16} & $\times$ &  & $\bullet$ & & $\bullet$ & & &  $\bullet$ &  $\pie{270}$ &  $\pie{90}$ & sym. $+$ SC & Server gets a key from & Server$/$Client tokenises & Server$/$Client encrypts & Exact $\&$ multi$-$rules \\
				&  &  &  & &  & & &  &  &  &  & admin at client, admin & messages & tokens & match: encrypted tokens \\
				&  &  &  & &  & & &  &  &  &  & encrypts rules, passes to MB & & & vs rules \\\hline
 				BlindIDS~\cite{CanardDKPS17}			&  $\checkmark^\$$ &  & $\bullet$ & & $\bullet$ & & & &  $\pie{270}$ &  $\pie{90}$ & Pairing & Server gen. key pair & Client tokenises messages & Client encrypts tokens & Exact match:\\
 				&  &  &  & &  & & & &  &  &  & Rule gen$/$editor gen. encrypted &  & using server public key  & encrypted tokens vs rules\\ 
 				&  &  &  & &  & & & &  &  &  & rules, passes them to MBs &  &  &\\\hline
				SPABox~\cite{FanGRCQ17}$^a$ &  $\times$ &  & $\bullet$ & & $\bullet$ & & & & $\pie{270}$ & $\pie{90}$ & DLP $+$ HE & Rule generator prepares & Client tokenises messages & Client encrypts tokens & Exact match, regular\\
				&  &  &  & &  & & & &  &  &  & rules & Client $\&$ Server negotiates &  & expressions, ML: Encrypted \\
				&  &  &  & &  & & & &  &  &  &  & DLP/HE parameters &  & tokens vs rules/models\\\hline
				PrivDPI~\cite{NingPLCC19}$^a$			&  $\times$ &  & $\bullet$ & & $\bullet$ & & & &  \pie{270} &  \pie{90} & sym. $+$ DLP & Rule generator prepares $\&$ & Client tokenises messages & Client encrypts tokens & Exact match:\\
				&   &  &  & &  & & & &  &  &  & signs rules & Client/MB/Server jointly &  & encrypted tokens vs rules\\
				&   &  &  & &  & & & &  &  &  &  & gen. reusable encrypted rules &  &\\\hline
				Pine~\cite{NingHPXLWD20}$^a$			&  $\times$ &  & $\bullet$ & & $\bullet$ & & & &  $\pie{270}$ &  $\pie{90}$ & sym. $+$ DLP & Enterprise gateway shares a key & Gateway tokenises messages  & Client encrypts tokens & Exact match:\\
				&  &  &  & &  & & & &  &  &  & with rule generator, rule & from client, gateway$/$MB jointly &  & encrypted tokens vs rules\\
				&  &  &  & &  & & & &  &  &  & generator prepares $\&$ signs rules & gen. reusable encrypted rules &  &\\\hline
				EV-DPI~\cite{ren2020privacy}			&  $\times$ &  & $\bullet$ & & $\bullet$ & & & $\bullet$ &  $\pie{270}$ &  $\pie{90}$ & sym. $+$ BF  & Enterprise gateway encrypts & Gateway tokenises messages & Gateway encrypts & Filter $\&$ exact match:\\
				&  &  &  & &  & & & &  &  &  $+$ CH & rules, passes encrypted & from client & tokens & encrypted tokens vs rules\\
				&  &  &  & &  & & & &  &  &  & rules to MBs & & &  \\\hline
				Yamada {\em et\ al.}~\cite{YamadaMTSP07}			&  $\times$ &  $\bullet$ &  & &  & $\bullet$ & & & $\CIRCLE$ & $\pie{180}$ & Statistical & Features: data size, time, HTTP & Feature vector extraction & n$/$a & Anomaly detection for IDS:\\
				&  &  &  & &  &  & & &  &  & analysis & traffic, access frequencies &  &  & Frequency analysis\\\hline
				Anderson {\em et\ al.}~\cite{AndersonM16,AndersonM17,AndersonPM18} &  $\times$ &  $\bullet$ &  & &  & $\bullet$ & & & $\CIRCLE$ & $\pie{180}$ & ML & Features: flow metadata, & Feature vector extraction & n$/$a & Malware traffic\\
				 & & &  & &  &  & & & & & classifiers & TLS handshake msg. etc. & & & classification\\\hline
				SGX-Box~\cite{HanKHH17}					&  $\times$ &  &  $\bullet$ & &  & & $\bullet$ & $\bullet$ & $\CIRCLE$ & $\pie{270}$ & SGX & Configure SGX, attest module, & Server securely shares & Enclave decrypts & Enclave inspects \\
				&  &  &   & &  & &  &  &  &  &  & $\&$ update server app. & session key with enclave & messages & decrypted payload\\\hline
				EndBox~\cite{GoltzscheRN+18}		&  $\times$ &  &  $\bullet$ & &  & & $\bullet$ & Edge & $\CIRCLE$ & $\pie{270}$  & SGX & Configure SGX, attest module, & Client app. securely shares  & Enclave at client & Enclave at client \\
				&  &  &  & &  & &  &  &  &   &  & enclave installs CA cert, & session key with enclave at & decrypts messages & inspects payload\\
				&  &  &  & &  & &  &  &  &   &  & enclave gen. key pair &  client (e.g. Edge device) &  &\\\hline
				SafeBricks~\cite{PoddarLPR18}				&  $\times$ &  &  $\bullet$ & &  &  & $\bullet$ &  $\bullet$ & $\CIRCLE$ & \pie{270}  & SGX & Configure SGX, attest module, & Enterprise gateway decrypts & Enclave at cloud & Enclave at cloud \\
				&   &  &   & &  &  &  &   &  &   &  & embed network functions & TLS traffic from clients, tunnels & decrypt messages & inspects payload\\
				&   &  &   & &  &  &  &   &  &   &  & in the enclave & to cloud enclave via IPSec &  &\\\hline
			\end{tabular}
			\begin{tablenotes}\footnotesize
				\item[] ``chg. TLS": Required changes to TLS, Pri.: Primitives, e.g. crypto primitives or metadata, Cl: Cloud, L: information leakage Util.: Utility, Pa.: Passive, Act.: Active, AC: Access Control, SE: Searchable Encryption, ML: Machine Learning, TH: Trusted Hardware, sym.: symmetric primitives for encryption$/$decryption (i.e. AES), SMC: Secure Multi-party computation, SC: Secret sharing, DLP: Discrete Log Problem, HE: Homomorphic encryption, SGX: Intel implementation of secure enclave, BF: Bloom Filter, CH: Cuckoo Hashing. 
				\item[] $a$: SpaBox, PrivDPI and Pine focus on improving the performance of BlindBox with added properties. SpaBox enables more expressive matching. PrivDPI introduces reusable obfuscated rules so that obfuscated rules can be re-use in many sessions, as compared to BlindBox that requires re-encryption of rules for every session. Pine introduces rule-hiding so that MBs do not learn the rules, and dynamic addition of rules.
				\item[] $\$$: BlindIDS proposes a new protocol based on pairing, that can be deployed as a replacement to SSL/TLS
				\item[] $*$: parts of the underlying encrypted traffic are decrypted based on access rights. This means MB sees part of the plaintext.
				\item[] $@$: Similar configuration with mbTLS but the protocol constructed is a modified version of the TLS protocol. In contrast, mbTLS only requires the use of TLS extension, and hence does not affect existing TLS implementation.
				\item[] $+$: fix security issues in mcTLS.
				\item[] $\#$: can be instantiated with unmodified TLS 1.3 draft 23. Constructed a provably secure protocol. The protocol is not meant to encourage adoption of active proxying but to demonstrate the difficulty of constructing a secure proxied end-to-end security protocol 
			\end{tablenotes}
	\end{threeparttable}}
	\end{adjustwidth}
\end{table*}
\end{landscape}
  \clearpage
}

%
%


\subsection{Performance}

\paragraph{\bf SE-based schemes preserve privacy but incur huge overhead.}
Schemes based on SE technique require two communication channels (e.g. one for the TLS connection and one for the encrypted tokens), the generation of tokens and encryption of rulesets (cf. Section~\ref{subsec:SE}).
This means SE-based schemes incur extra overheads, at least compare to the MitM approach.
However, it represents a promising technique to privately inspect encrypted payloads without needing any specialised hardware as in the TH-based solutions.
It also does not reveal partially the underlying data, as in the case of AC-based solutions, except when regular expression type of matching is required.
The challenge is then to improve the efficiency on the current schemes especially in terms of rulesets and encryption and matching.
A question that need to be addressed is that in the case where specialised hardware is widely-available, would SE-based approach still play a role in providing inspection of encrypted traffic in a private manner?
The answer would be yes if combining the SE-based approach and the TH approach to lessen the processing loads of TH is relatively more efficient than having all encrypted traffic being processed in the trusted hardware. 

\paragraph{\bf ML approach presents ideal solution but inefficient?} 
Schemes based on ML represents the ideal solution since existing setup is not required to be changed.
However, performance based on ML matching may need to be further explored especially when compared with all the other approaches.

\paragraph{\bf AC-based schemes as a replacement to MitM?}
In order to avoid the security issues of the MitM-based solutions, schemes based on AC technique proposes protocol that requires all MBs to be accountable between the client and the server.
If this is deployed it means the client and the server are able to setup a secure session with each of the MB and decide what data the MB is allowed to view.
The performance would then be similar to a plain MitM-based solution since the MB performs decryption, inspection and re-encryption as before, but in the AC setting.
The challenge is not so much on performance in this case, but security, configuration and utility, which we discuss in the other two sections.

\paragraph{\bf Moving towards TH-based solutions.}
As per our observation on the most recent literatures, new schemes are leaning towards utilizing trusted hardware efficiently such as the secure enclave technology provided by Intel SGX that is available in most of the recent Intel processors.
The challenge is the continuous efforts in improving the performance while at the same time minimize leakages of information.
This is crucial since the secure enclave may be considered resource-constrained device in a certain sense, and not all network traffic should be routed into the memory or storage space of the enclave for processing.
The research direction is thus to examine and construct efficient, yet private communication protocol between the enclave and the MB (and/or the service provider).
Alternatively, a scheme may construct a secret share protocol that utilize multiple enclaves that distribute workload between these enclaves in a secure manner.

\subsection{Utility}
In terms of utility, schemes based on AC and TH have the capability to provide full functionality similar to inspection on plaintext data.
This is because both techniques enable MBs to decrypt the underlying traffic.

\paragraph{\bf Full functionalities without client-server accountability.}
The challenge for schemes based on AC technique is that whether it is possible to achieve such utility without having to a priori decide and authenticate the MBs involved in the communication between the client and the server.
It is not clear when a new MB is introduced, or an existing one being removed, how the client and the server update their communication.
Furthermore, AC introduces the concept of context, where the client and the server have the flexibility of setting a MB access policy towards their data.
This means deciding the MB that can read and/or write a particular section of the encrypted data.
It is also not clear how this can be performed in a systematic and accurate manner, especially there are information in different domain that may requires specific access policy.
As was stated in~\cite{Carnavalet20}, solutions based on AC technique require the support of the server and the client, which may not be feasible since for example the server has no interest to help a client that would like to prevent malware download.

\paragraph{\bf Extending functionalities of SE and ML Techniques.}
In the case where deploying specialised hardware is not an option, especially with legacy system, one may seek to extend the limited functionalities of the SE-based and ML-based approach.
The difficulty for SE-technique is the complications of extending the token matching mechanism without leaking substantial information.
While Embark~\cite{LanSPRL16} and Yuan {\em et\ al.}~\cite{YuanWLW16} extended the capability of BlindBox~\cite{SherryLPR15}, the matching still does not provide full regular expression matching.
For ML-based schemes, the challenge would be to construct ML algorithms and models that enable detection of different-type of anomaly traffics without inspecting the payload.

%
%


%
%
\section{Conclusions}
\label{sec:concl}
In this work we present a comprehensive survey on the topic of privacy-preserving inspection over encrypted traffic.
We define a trust model and categorise the different network settings based on existing state-of-the-art schemes.
From our compilation, we further categorise the current schemes into four main techniques, that enable us to demonstrate the advantages and limitations of each of the proposals.
These gave insights into the suitability of the proposals to be deployed in practice, which we also discussed.
The main difficulty is to fill the gap between actual deployment, which very much still based on the man-in-the-middle approach that does not preserve privacy.
To this, we listed and discussed the many challenges faced in the existing techniques and possible directions for improvements.

\begin{acks}
Research supported by the National Research Foundation,
Prime Minister’s Office, Singapore, under its Corporate Laboratory@
University Scheme, National University of Singapore, and
Singapore Telecommunications Ltd.
\end{acks}

\bibliographystyle{ACM-Reference-Format}
\bibliography{dpi}




\end{document}